\documentstyle[12pt,epsf]{elsart}

\begin{document}

\begin{frontmatter}

\title{EXPECTED  BEHAVIOUR OF DIFFERENT  SEMICONDUCTOR  MATERIALS  IN  
HADRON  FIELDS}

\author[univ]{I.Lazanu}, 
\author[iftm]{S.Lazanu} and
\author[infn]{M.Bruzzi} 

\address[univ]{University of Bucharest, Faculty of Physics,
P.O.Box  MG-11, Bucharest-Magurele, Romania,
electronic address: ilaz@scut.fizica.unibuc.ro}
\address[iftm]{National Institute for Materials Physics,
P.O.Box MG-7, Bucharest-Magurele, Romania, 
electronic address: lazanu@alpha1.infim.ro}
\address[infn]{Universita di Firenze, Dipartimento di Energetica,
Via S. Marta 3, 50139 Firenze, Italy,
electronic address: bruzzi@fi.infn.it}

\begin{abstract}
The utilisation of semiconductor materials as detectors and devices 
operating in high radiation environments, at the future particle 
colliders, in space applications or in medicine and industry, 
necessitates to obtain radiation harder materials. A systematic 
theoretical study has been performed, investigating the interaction of 
charged hadrons with semiconductor materials and the mechanisms of 
defect creation by irradiation. The mechanisms of the primary 
interaction of the hadron with the nucleus of the semiconductor lattice 
have been explicitly modelled and the Lindhard theory of the partition 
between ionisation and displacements has been considered. The behaviour 
of  silicon, diamond, and some A$^{III}$B$^V$ compounds, as GaAs, GaP, InP, InAs, 
InSb has been investigated. The nuclear energy loss, and the 
concentration of primary defects induced in the material bulk by the 
unit hadron fluence have been calculated. The peculiarities of the 
proton and pion interactions as well as the specific properties of the 
semiconductor material have been put in evidence.
\medskip

61.82.-d: Radiation effects on specific materials.\\
\medskip
\end{abstract}

\end{frontmatter}
\section{Introduction}

The crystalline materials for semiconductor devices used in high 
fluences of particles are strongly affected by the effects of radiation. 
After the interaction between the incoming particle and the target, 
mainly two classes of degradation effects are observed: surface and bulk 
material damage, due to the displacement of atoms from their sites in the
lattice. For electrons and gammas the effects are dominantly at the 
surface, while the heavy particles (pions, protons, neutrons, ions) 
produce both types of damages.

Up to now, in spite of the experimental and theoretical efforts, 
the problems related to the behaviour of the semiconductor materials in 
radiation fields, the identification of the induced defects and their 
characterisation, as well as the explanation of the degradation 
mechanisms are still open problems.

The utilisation of semiconductor materials as detectors and devices 
operating in high radiation environments, at the future particle 
colliders, in space applications, in medicine and industry, makes 
necessary to obtain radiation harder materials. 

The diamond and different A$^{III}$B$^V$ compounds (GaAs, GaP, InP, InAs, InSb) 
are in principle, possible competitors for silicon to different 
electronic devices. In the present paper a systematic theoretical study 
of the expected behaviour of these semiconductor materials in hadron 
fields (protons and pions) is presented.

All analysed materials have a zinc-blend crystalline structure. 
Silicon is at the base of electronic industry, diamond and the A$^{III}$B$^V$ 
compounds present attractive electrical and/or luminescence properties 
\cite{1}, of interest for different applications.

\section{Interaction of charged hadrons with semiconductor materials and
mechanisms of defect creation by irradition}

The charged particles interact with both atomic and electronic systems 
in a solid. The total rate of energy loss, as a function of distance, 
is called the stopping power. This one could, in general, be divided 
artificially into two components, the nuclear and the electronic part:

$\left( -\frac{dE}{dx}\right) _{total}=\left( -\frac{dE}{dx}\right)
_{nuclear}+\left( -\frac{dE}{dx}\right) _{electronic}$

Roughly speaking, the energy lost due to interactions with electrons 
gives rise to material ionisation, while the energy lost in interactions 
with nuclei is to the origin of  defect creation.

A comprehensive theoretical treatment of electronic stopping which 
covers all energies of interest cannot be formulated simply because of 
different approximations concerning both the scattering and contribution 
of different electrons in the solid. For fast particles with velocities 
higher then the orbital velocities of the electron, the Bethe-Bloch 
formula is to be used \cite{2}. At lower velocities, inner electrons have 
velocities greater than particle velocity, and therefore do not 
contribute to the energy loss. This regime was modelled by Lindhard and 
Scharff \cite{3}.

If the particle has a positive charge, and a velocity close to the 
orbital velocity of its outer electrons, it has a high probability of 
capturing an electron from one of the atoms of the medium through which 
it passes. This process contributes to the total inelastic energy loss 
since the moving ion has to expend energy in the removal of the electrons 
which it captures.

The nuclear stopping depends on the detailed nature of the atomic 
scattering, and this in turn depends intimately on the form of the 
interaction potential. At low energies, a realistic potential based on 
the Thomas-Fermi approximation was used in the literature \cite{3} and at 
higher energies, where scattering results from the interaction of 
unscreened nuclei, a Rutherford collision model is to be used.

In Figure 1, the nuclear stopping power calculated for protons and pions 
in diamond, silicon and GaAs is represented, as a function of the kinetic 
energy of the particle. The nuclear stopping power is greater for heavier 
incident particles (protons compared with pions), and for lighter media 
(diamond in comparison with silicon and GaAs). The position of its 
maximum is the same for protons and pions in the same medium.

The process of partitioning the energy of the recoil nuclei (produced 
due the interaction of the incident particle with the nucleus, placed in 
its lattice site) by new interaction processes, between electrons 
(ionisation) and atomic motion (displacements) is considered in the 
frame of Lindhard theory \cite{4}.

The mechanism considered in the study of the interaction between the 
incoming particle and the solid, by which bulk defects are produced, is 
the following: the particle, heavier than the electron, with electrical 
charge or not, interacts with the electrons and with the nuclei of the 
crystalline lattice. The nuclear interaction produces the bulk defects. 
As a result of the interaction, depending on the energy and on the 
nature of the incident particle, one or more light particles are formed, 
and usually one or more heavy recoil nuclei. These nuclei have charge 
and mass numbers lower or at least equal with the medium. After this 
interaction process, the recoil nucleus or nuclei, if they have 
sufficient energy, are displaced from the lattice positions in 
interstitials. Thus, the primary knock-on nucleus, if its energy is 
large enough, can produce the displacement of a new nucleus, and the 
process can continue as a cascade, until the energy of the nucleus 
becomes lower than the threshold for atomic displacements. Because of 
the regular nature of the crystalline lattice, the displacement energy 
is anisotropic. In the present model, averaged values for displacement 
energies have been considered. In the concrete evaluation of defect 
production, the nuclear interactions must be modelled, see for example 
\cite{5,6,7}. The primary interaction between the hadron and the nucleus of 
the lattice presents characteristics reflecting the peculiarities of the 
hadron, especially at relatively low energies. If the inelastic process 
is initiated by nucleons, the identity of the incoming projectile is 
lost, and the creation of  secondary particles is associated with energy 
exchanges which are of the order of MeV or larger. For pion nucleus 
processes, the absorption, the process by which the pion disappears as a 
real particle, is also possible.

The energy dependence of cross sections, for proton and pion interaction 
with the nucleus, present very different behaviours: the proton-nucleus 
cross sections decrease with the increase of the projectile energy, then 
have a minimum at relatively low energies, followed by a smooth increase, 
while the pion nucleus cross sections present for all processes a large 
maximum, at about 160 MeV, reflecting the resonant structure of 
interaction (the $\Delta _{33}$ resonance production), followed by other resonances, 
at higher energies, but with much less importance. Due to the multitude 
of open channels in these processes, some simplifying hypothesis have 
been done \cite{8}.

The physical quantity that characterise the primary defects is the 
concentration of primary radiation induced defects on the unit particle 
fluence (CPD) \cite{6}. This quantity permits the correlation of damages 
produced in different materials at the same kinetic energy of the 
incident hadron. For the comparison of the effects of different 
particles in the same semiconductor material, the non ionising energy 
loss (NIEL) is useful.

\section{Theoretical expected material behaviour} 

The behaviour of these materials in proton fields is characterised by 
the CPD. In Figure 2, the dependence of the CPD as a function of the 
protons kinetic energy and medium mass number (diamond, silicon, GaAs 
and InP) is presented - see references \cite{5,9} and references cited 
therein. Low kinetic energy protons produce higher degradation in all 
materials.

For pion induced degradation, the energy dependence of CPD and NIEL 
presents two maxima, the relative importance of which depends on the 
target mass number: one in the region of the $\Delta _{33}$ resonance, more 
pronounced for light elements and compounds containing light elements, 
and another one around 1 GeV kinetic energy, more pronounced for heavy 
elements. At higher energies, an weak energy dependence is observed, and 
a general  dependence of the NIEL can be approximated \cite{8,10}. In 
Figure 3a, the CPD for all these materials is represented as a function 
of the pion kinetic energy and of material average mass number.

These materials could be separated into two classes, the first 
with monoatomic materials or materials with relatively close mass 
numbers, and the second comprising binary materials with remote mass 
numbers of the elements. In Figures 3b and 3c  
respectively, the energy and mass number dependence of the CPD for the 
two groups of materials are represented separately.

A slow variation of the primary defect concentration has been found for 
pion irradiation of diamond, silicon, GaP and GaAs, in the whole energy 
range of interest, with less than 2 displacements/cm/unit of fluence. In 
contrast to this situation, GaP, InAs and InSb are characterised by a 
low CPD in the energy range up to 200 MeV, followed by a pronounced 
increase of displacement concentration, to more than 8 
displacements/cm/unit of fluence at high energies.

\section{Summary}

A systematic theoretical study has been performed, investigating the 
interaction of charged hadrons with semiconductor materials and the 
mechanisms of defect creation by irradiation. The mechanisms of the 
primary interaction of the hadron with the semiconductor nucleus lattice 
have been explicitly modelled and the Lindhard theory of the partition 
between ionisation and displacements has been considered.

The nuclear stopping power is greater for heavier incident particles 
(protons compared with pions), and for lighter media (diamond in 
comparison with silicon and GaAs). The position of its maximum is the 
same for protons and pions in the same medium.

The behaviour of these materials in hadron fields is characterised by 
the CPD and NIEL.

For protons, the low kinetic energy protons produce higher degradation 
in all materials.

For pion induced degradation, the energy dependence of CPD and NIEL 
presents two maxima, the relative importance of which depends on the 
target mass number: one in the region of the $\Delta _{33}$ resonance, more 
pronounced for light elements and compounds containing light elements, 
and another one around 1 GeV kinetic energy, more pronounced for heavy 
elements. At higher energies, an weak energy dependence is observed. A 
slow variation of the primary defect concentration has been found for 
pion irradiation of diamond, silicon, GaP and GaAs, in the whole energy 
range of interest, with less than 2 displacements/cm/unit of fluence. 
In contrast to this situation, GaP, InAs and InSb are characterised by a 
low CPD in the energy range up to 200 MeV, (which represent the energy 
range up to their utilisation in pion field is recommended), followed by 
a pronounced increase of displacement concentration to more than 8 
displacements/cm/unit of fluence at high energies.

\newpage
\begin{center}
\bf{Figure captions}
\end{center}
\bigskip
\medskip

Figure 1:
Nuclear stopping power of protons and pions in diamond, silicon and GaAs.
\medskip

Figure 2:
The concentration of primary induced defects for unit of fluence (CPD) 
in diamond, silicon, GaAs and InP by protons. The mesh surfaces are 
drawn only to guide the eye only. The discontinuity in the surface 
represent the behaviour. The data for proton - diamond are from \cite{5}, 
the silicon data are averaged values from \cite{11} and  \cite{12}, and that for 
GaAs and InP are from \cite{12}.

\medskip
Figure 3a:
The energy and material dependence of the concentration of primary defects
on unit pion fluence in diamond, silicon, GaP, GaAs, InP, InAs and InSb.
The mesh surfaces are plotted for guide the eyes. The differences in the
behaviour of these materials are clearly suggested by the discontinuity in
the mesh surface.

\medskip
Figure 3b:
The energy and material dependence of the concentration of 
primary defects on unit pion fluence, induced by pions in diamond, 
silicon, GaAs and InSb. The mesh surfaces are drawn to guide the eyes. 
A pronounced maximum, corresponding to the $\Delta _{33}$ resonance is clearly 
visible, and the second maximum at higher energies is important 
especially for InSb.

\medskip
Figure 3c:
The same as in Figure 3b, but for GaP, InP and InAs. The effects of the
$\Delta _{33}$ resonance are less pronounced
in respect to the effects of the inelastic region, at high energies.

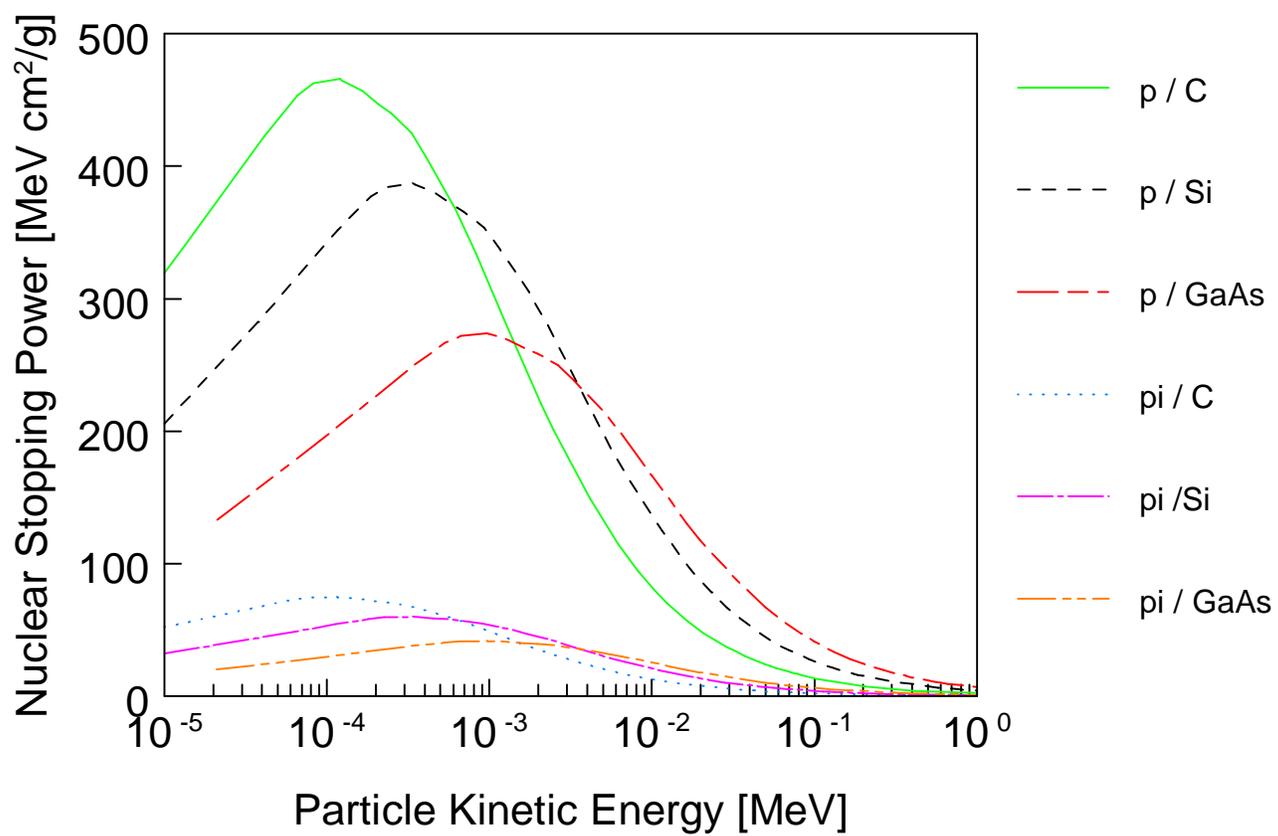
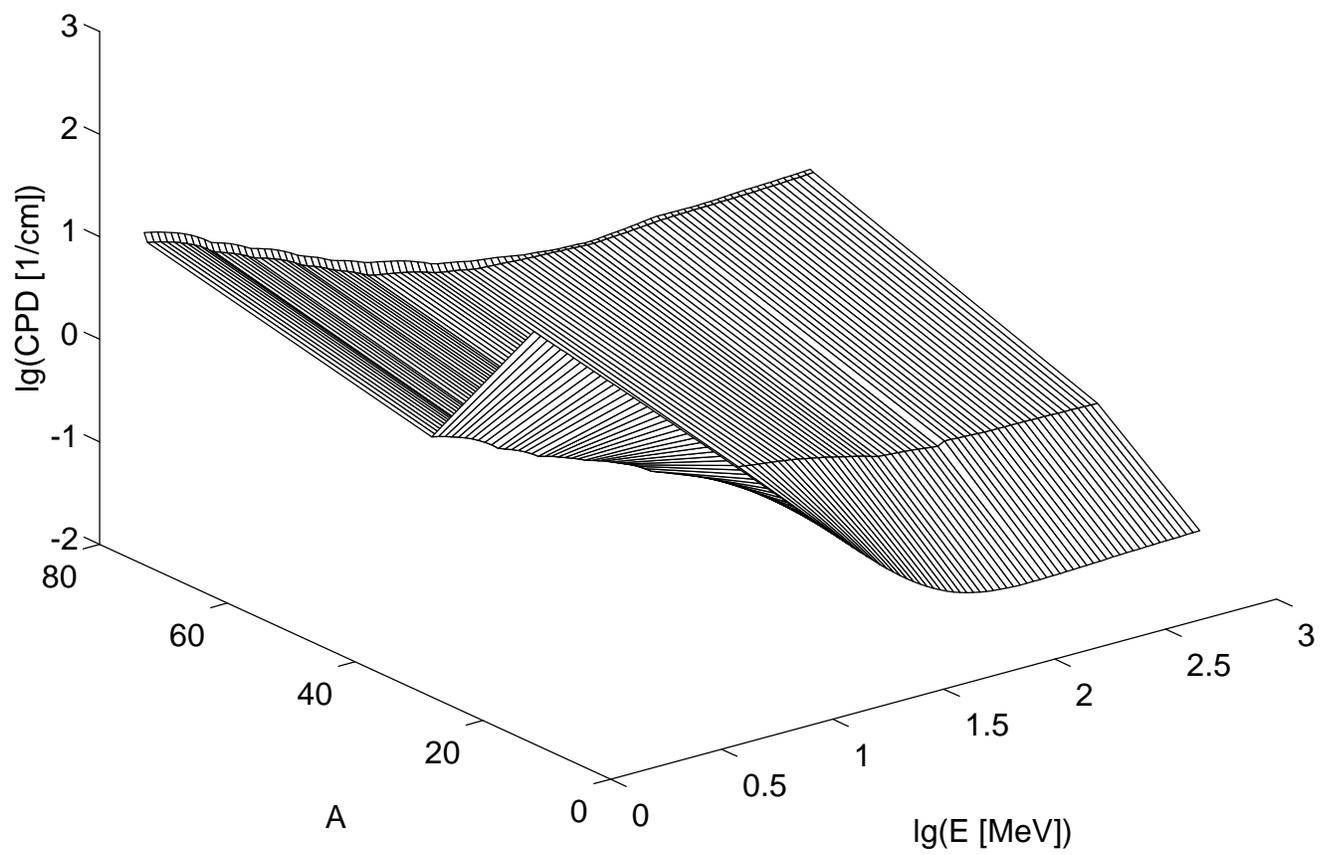
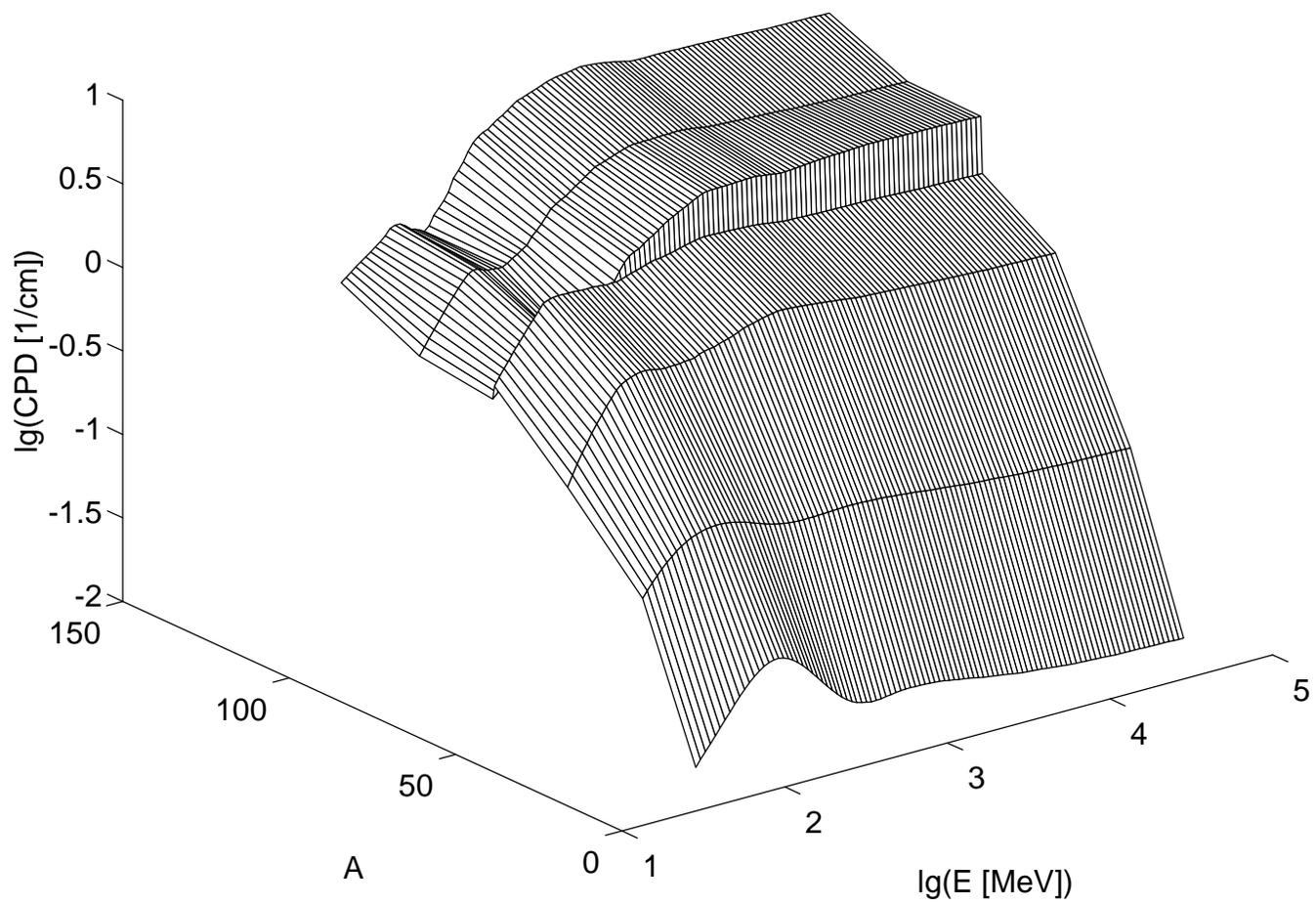
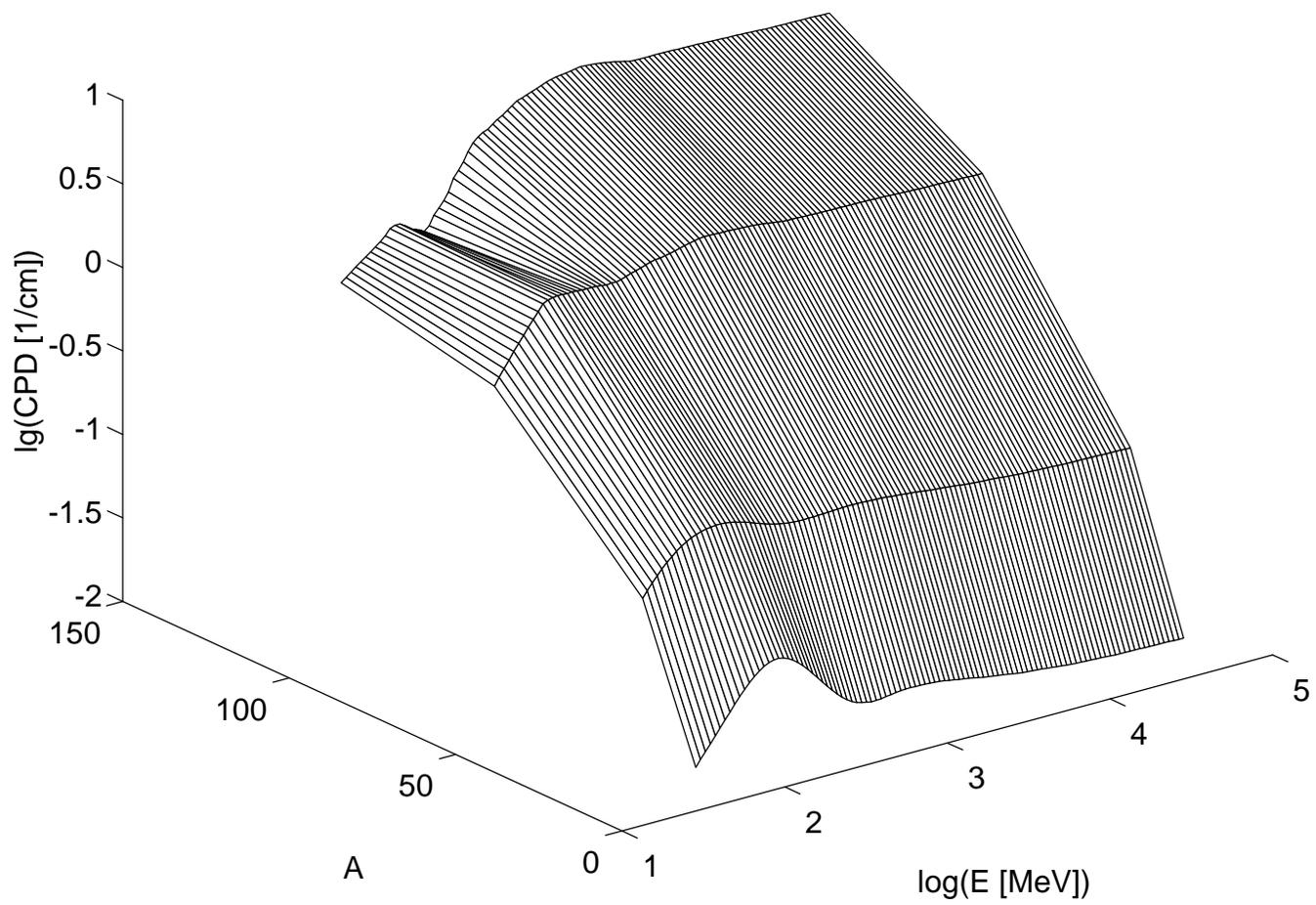
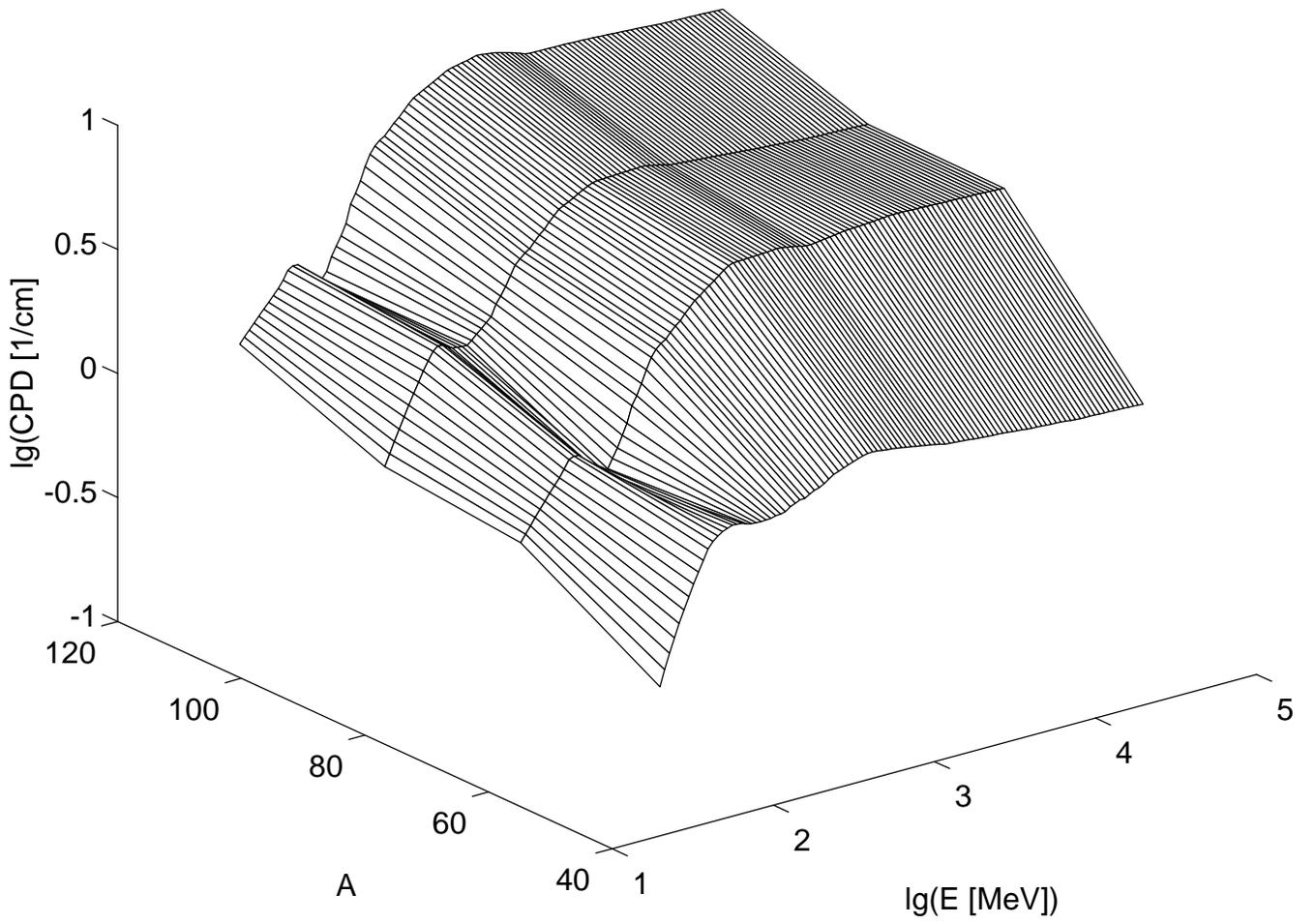
\end{document}